\documentclass[10pt,conference,letterpaper]{IEEEtran}

\usepackage[utf8]{inputenc}
\usepackage{graphicx}
\usepackage{amssymb}
\usepackage[cmex10]{amsmath}
\usepackage{color}
\usepackage{theorem}
\usepackage{pifont}
\usepackage[ruled]{algorithm2e}

\usepackage{bbm}

\renewcommand{\mod}{ \text{\, mod \,} }

\newcommand{\rpolx}[1]{\ensuremath{r^{(#1)}(x)}}

\newcommand{\qpolx}[1]{\ensuremath{q^{(#1)}(x)}}
\newcommand{\qpol}[1]{\ensuremath{q^{(#1)}}}
\newcommand{\qhpolx}[1]{\ensuremath{\hat{q}^{(#1)}(x)}}

\newcommand{\qbpolx}[1]{\ensuremath{\bar{q}^{(#1)}(x)}}
\newcommand{\qbpol}[1]{\ensuremath{\bar{q}^{(#1)}}}
\newcommand{\upolx}[1]{\ensuremath{u^{(#1)}(x)}}
\newcommand{\upol}[1]{\ensuremath{u^{(#1)}}}
\newcommand{\vpolx}[1]{\ensuremath{v^{(#1)}(x)}}
\newcommand{\vpol}[1]{\ensuremath{v^{(#1)}}}

\newcommand{\DPx}[1]{\ensuremath{\Delta^{(#1)}(x)}}
\newcommand{\DPol}[1]{\ensuremath{\Delta^{(#1)}}}

\newcommand{\cval}[1]{\ensuremath{c^{(#1)}}}
\newcommand{\da}{\ensuremath{\DPol{1}(\alpha_1)}}
\newcommand{\daa}{\ensuremath{\DPol{1}(\alpha_2)}}
\newcommand{\dda}{\ensuremath{\DPol{2}(\alpha_1)}}
\newcommand{\ddaa}{\ensuremath{\DPol{2}(\alpha_2)}}

\newcommand{\mfloor}[1]{\ensuremath{\left\lfloor #1 \right \rfloor}}
\newcommand{\dhalf}{\ensuremath{\mfloor{\frac{d-1}{2}}}}
\newcommand{\reff}[1]{(\ref{#1})}
\newcommand{\order}{\ensuremath{\mathcal O}}

\newcommand{\mx}{\mathcal{X}}
\newtheorem{problem}{Problem}

\renewcommand{\deg}{\textnormal{deg}\ }

\IEEEoverridecommandlockouts 

\begin{document}

\title{The Euclidean Algorithm for Generalized Minimum Distance Decoding of Reed-Solomon Codes}

\author{\IEEEauthorblockN{Sabine Kampf and Martin Bossert}
\IEEEauthorblockA{Institute of Telecommunications and Applied Information Theory\\
University of Ulm, Germany\\
\texttt{\{sabine.kampf | martin.bossert\}@uni-ulm.de}}
\thanks{This work was supported by the German Research Council "Deutsche Forschungsgemeinschaft" (DFG) under Grant No. Bo867/22.}}

\maketitle

\begin{abstract}
This paper presents a method to merge Generalized Minimum Distance decoding of Reed-Solomon codes with the extended Euclidean algorithm. By merge, we mean that the steps taken to perform the Generalized Minimum Distance decoding are similar to those performed by the extended Euclidean algorithm. The resulting algorithm has a complexity of $\order(n^2)$.
\end{abstract}


\section{Introduction}
In 1996, Ralf K\"otter presented a fast algorithm for Generalized Minimum Distance (GMD) decoding of Reed-Solomon (RS) codes with a complexity of $\order(n^2)$ \cite{Koetter_GMD}. This algorithm is an extension of the well-known Berlekamp-Massey algorithm (BMA), that has been applied to decoding of RS codes up to half the minimum distance since the late 1960s \cite{MacWilliams}. Another algorithm that is often used to decode RS codes was first described by Sugiyama et. al. in 1975 \cite{Sugiyama}, and is based on the extended Euclidean algorithm (EEA). Recently, there have been attempts to perform GMD decoding, also with a complexity of $\order(n^2)$, using polynomials that are obtained from the extended Euclidean algorithm \cite{scc_wir}, \cite{isit2010}. Since the BMA and EEA are known to be equivalent in the decoding of RS codes up to half the minimum distance, it is interesting to try to find an extension to the EEA that is equivalent to K\"otters extension to the BMA. In this paper, we present such an extension.

The paper is organized as follows: In the next section, we shortly introduce RS codes as well as GMD decoding. In Section \ref{sec:EEA}, we first recall the EEA and how it is used for decoding RS codes up to half the minimum distance. Then we show, which of polynomials obtained are used as a basis for the GMD extension, and how to extend the EEA to GMD decoding. Finally, we shortly discuss how to modify our approach so that the selection of the best solution does not decrease the complexity. The conclusion follows in Section \ref{sec:summary}.

\section{Notations and Definitions}
\subsection{RS Codes and Key Equation} \label{ssec:RSKE}
To define an $\mathcal{RS} (n, k, d)$ code over $GF(q)$ with $R = \frac{k}{n}$ and minimum distance $d=n-k+1$, let $\alpha \in GF(q)$ denote an element of order $n$, and $\alpha^j$ the $j$th power of this element. Later, the elements $\alpha_i$ will denote the $i$th element according to a (implicitly) given indexing. A polynomial $c(x)=c_0+c_1x+\dots+c_{n-1}x^{n-1}$ with coefficients in $GF(q)$ is a valid codeword if the polynomial $C(x)$, whose coefficients $C_j$ are calculated by the discrete Fourier transform (DFT), i.e.
\begin{equation}
C_j = c(\alpha^{j}), \qquad \, j=0,\dots,n-1,
\end{equation}
is zero at the first $n-k=d-1$ coefficients, hence 
\begin{equation}
C(x)=C_{d-1}x^{d-1}+\dots+C_{n-1}x^{n-1}.
\label{eqn:cwdef}
\end{equation}
We take the $C_i \in GF(q)$ to be the information symbols. The codeword $c(x)$ corresponding to the information word $C(x)$ is obtained through the inverse discrete Fourier transform (IDFT):
\begin{equation}
c_i = n^{-1}\cdot C(\alpha^{-i}),\quad i=0,\dots,n-1.
\end{equation}
Throughout this paper, capital letters denote polynomials in the spectral domain, and small letters their correspondences in the time domain.

The transmitted codeword is corrupted by an additive error $e(x)$ of Hamming weight $t$, and the received word is $r(x)=c(x)+e(x)$. For decoding, calculate the syndrome $S(x)$:
\begin{equation}
S(x) = R(x) \mod x^{d-1} = E(x) \mod x^{d-1}.
\end{equation}
This syndrome is used in the \textit{key equation} for decoding RS codes:
\begin{equation} \label{eqn:key_eq}
-\Omega(x) \equiv \Lambda(x)\cdot S(x)  \mod x^{d-1},
\end{equation}
with the error locator polynomial $\Lambda(x)$ and the error evaluator polynomial $\Omega(x)$. These two polynomials satisfy the important degree relation:
\begin{equation}\label{eqn:deg_omega}
\deg\Omega(x) < \deg \Lambda(x) =t.
\end{equation}
\subsection{GMD Decoding} \label{ssec:GMD}
First proposed by Forney in 1966 \cite{Forney_GMD}, the idea of GMD decoding is to allow soft-decision decoding of algebraically decodable codes by performing $m+1$ decoding trials, each with a different number of positions being erased. In each trial, the decoder may either output an error locator polynomial as defined in \eqref{eqn:key_eq}, or declare a decoding failure. Consequently, a list of up to $m+1$ candidate error locator polynomials is obtained, and the decoder should select one of these candidate words, that is best according to a certain criterion. 

In order to determine the set $\mx_j$ of erased positions in iteration $j$, the decoder needs to be supplied with reliability information regarding the decisions made. Reliability is defined intuitively, i.e. the less reliable a decision, the more probable it is that the received value is incorrect. Therefore, the least reliable positions are erased first. It is known, cf. e.g. \cite{Koetter_GMD}, that decoding is possible if the number of errors $t$ and the number of erasures $\epsilon$ fulfill 
\begin{equation}
2t+\epsilon < d.
\label{eqn:2tpeltd}
\end{equation}
We choose $|\mx_0|=0$ (i.e. we start with decoding errors only), $|\mx_1|=2$, $|\mx_2|=4$, \dots, $|\mx_{m}|=2\cdot \dhalf$, and further require $\mx_j \subset \mx_{j+1}$. Before, $\Lambda(x)$ was defined to be an error locator polynomial. For our algorithm, we slightly modify this and take $\Lambda(x)$ to be a joint error-erasure locator polynomial. In each iteration of the GMD decoding process, we have two additional erasures. On the other hand, according to \eqref{eqn:2tpeltd} we can correct one error less than in the iteration before, and so we find that the degree of our candidate error locator polynomial should increase by $1$ in each iteration. Due to this fact, we state our decoding problem as follows: 
\begin{problem}
In each iteration of our GMD decoding, we want to find an error locator polynomial $\Lambda(x)$, that fulfills \eqref{eqn:key_eq} and \eqref{eqn:deg_omega}, of a \emph{given} degree with certain prescribed zeros. 
\end{problem}
This is different to the problem statement of K\"otter, where the polynomial of \emph{smallest} degree is to be found. It will be seen later that in some situations, it is not possible to fulfill the requirements given in our problem statement.

In ``classical'' GMD decoding, the iterations are independent of each other. In each iteration, one first determines the erasure locator polynomial. Due to the erasures, the minimum distance of the RS code is virtually decreased, and the decoder tries to find an error locator from the shorter syndrome. Because the complexity of decoding with the EEA 
is $\order(n^2)$, the overall decoding complexity of this approach is $\order(n^3)$. However, the decoding complexity can be decreased if erasing of positions is performed incrementally, i.e. the decoding result of iteration $j$ is used together with the additional erasures in $\mx_{j+1}\setminus\mx_{j}$ to yield the decoding result of iteration $j+1$. The first such method was presented in \cite{Koetter_GMD}, the complexity of his approach being $\order(n^2)$ (actually, K\"otter claims the complexity to be $\order(nd)$, but since in general $d=\order(n)$, this is asymptotically the same). His algorithm is an extension of the BMA. However, it is known that decoding up to $\dhalf$ with the BMA and the EEA are equivalent. Therefore, we want to show that it is possible to modify the EEA, such that GMD decoding is merged into the decoding process. 

\section{The Extended Euclidean Algorithm} \label{sec:EEA}
The possibility of applying the EEA in the decoding of RS codes up to $\dhalf$ was first presented by Sugiyama~et.~al. in 1975 \cite{Sugiyama}. This decoding approach will be shortly reviewed in the first part of this section, as it is the basis for the extended decoding approach. 

In the second and third part of this section, we will present an algorithm that integrates GMD decoding into the EEA. At the end of the third part, we show how to combine the formulas given to form the algorithm. 

\subsection{Decoding up to $\dhalf$} \label{ssec:EEAclass}
The EEA uses input polynomials $A(x)=\rpolx{0}$ and $B(x)=\rpolx{-1}$ to recursively calculate a series of quotient polynomials $\qpolx{j+1}$ and remainders $\rpolx{j+1}$ that fulfill:
\begin{equation}
\rpolx{j+1}=\rpolx{j-1} - \qpolx{j+1}\cdot\rpolx{j},
\label{eqn:EEAremrec}
\end{equation}
with $\deg \rpolx{j+1}<\deg \rpolx{j}$. The algorithm stops when $\rpolx{j+1}=0$. 
From the quotient polynomials, a series of auxiliary polynomials $\upolx{j+1}$ is obtained recursively, namely
\begin{equation}
\upolx{j+1}=\upolx{j-1} - \qpolx{j+1}\cdot\upolx{j},
\label{eqn:EEAauxrec}
\end{equation}
where $\upolx{-1}=0$ and $\upolx{0}=1$. The degrees of these auxiliary polynomials are given by
\begin{equation}
\deg \upolx{j} = \sum_{i=1}^{j} \deg \qpolx{i}.
\end{equation}
Further, these polynomials fulfill the relation 
\begin{equation}
\upolx{j}\cdot A(x) = \rpolx{j} \mod B(x),
\end{equation}
which has a form similar to the key equation \reff{eqn:key_eq}. This implies that the EEA can be used for solving \reff{eqn:key_eq}.
Hence by setting $A(x) = S(x)$ and $B(x)=x^{d-1}$, in some steps of the EEA, whenever 
\begin{equation}
\deg \upolx{j} > \deg \rpolx{j},
\label{eqn:EEAstopcond}
\end{equation}
we obtain polynomials fulfilling both \reff{eqn:key_eq} and \reff{eqn:deg_omega}.
If the number of errors $t$, i.e. the number of nonzero coefficients in $e(x)$, is limited by $t\leq \dhalf$, then it is known \cite{Sugiyama} that $t = \deg \upolx{j}$, $\Lambda(x) = \upolx{j}$ and $\Omega(x) = -\rpolx{j}$ where $j$ is the smallest index for which \reff{eqn:EEAstopcond} is fulfilled.

\subsection{From Classical Decoding to GMD Extension} \label{ssec:ubergang}
In \cite{isit2010} it was shown that $\cval{j+1}$, the number of coefficients of $\qpolx{j+1}$ that can be determined from the syndrome, is given by
\begin{equation}
\cval{j+1} = \deg \rpolx{j} - \deg \upolx{j} + 1.
\end{equation}
Further, \eqref{eqn:EEAremrec} yields that 
\begin{equation}
\deg \qpolx{j+1} = \deg \rpolx{j-1}-\deg \rpolx{j}.
\end{equation}
As long as $\deg \qpolx{j+1} + 1 \leq \cval{j+1}$, the quotient polynomials are calculated as in the classical decoding procedure. This can be done as long as $\deg \qpolx{j+1} \leq \dhalf$ \cite{isit2010}. If $\deg \qpolx{j+1} + 1 > \cval{j+1}$, we switch to the GMD extension described in the next paragraph. In order to set the initial polynomials for the extension, we have to distinguish two cases: If $\cval{j+1} \leq 0$, then no coefficient of the next quotient polynomial can be determined. Hence, we use 
\begin{equation}
\upolx{j}=:\DPx{1, 0} \ \textnormal{and} \ \upolx{j-1}=:\DPx{2, 0}.
\label{eqn:dpx0def}
\end{equation}
On the other hand, if $\cval{j+1}>0$, then it is possible to determine the upmost $\cval{j+1}$ coefficients of $\qpolx{j+1}$ - we will call this part $\qhpolx{1}$, because it belongs to the ``quotient'' polynomial determined in the first iteration of the GMD extension - and it would be unwise to discard this information. Simulations have shown, that the best performance is achieved if we set $\DPx{1, 0}:=\upolx{j}$ and $\DPx{2, 0}:=\upolx{j-1}-\qhpolx{1}\upolx{j}$. Unfortunately, we do not know yet why this performs better than using the definitions in \eqref{eqn:dpx0def} and just taking $\qhpolx{1}$ into account only in the first iteration.

\subsection{GMD Extension} \label{ssec:GMDext}
In this section, we again use $j$ to index the iterations done in the GMD part of our algorithm. However, we count those iterations independently of the ones performed by the EEA before. Because the decoding up to $\dhalf$ is just decoding without erasures, $j$ now corresponds to the iterations defined in Section \ref{ssec:GMD}. 
The basic idea is the following: We start with decoding up to half the minimum distance, as described before. Once the polynomials $\upolx{j}$ can no longer be determined by the syndrome and we switch to the GMD extension, the decoder starts to determine the ``quotient'' polynomials $\qpolx{j}$ with the help of the reliability information given. Namely, two positions $\alpha_1$ and  $\alpha_2$ are erased in each iteration, where 
\begin{equation}
\left\{\alpha_1, \alpha_2\right\}=\mx_{j+1}\setminus\mx_{j}.
\end{equation}
Comparing \eqref{eqn:dpx0def} to \eqref{eqn:EEAauxrec}, we set the equation to be solved by the GMD extension to be
\begin{equation}
\DPx{1, j+1}=\DPx{2, j}-\qpolx{j}\DPx{1, j}.
\label{eqn:EEAexteqn}
\end{equation}
Consequently, $\qpolx{j}$ is determined in such a way that 
\begin{equation}
\DPol{1, j+1}(\alpha_1)=\DPol{1, j+1}(\alpha_2)=0.
\end{equation}
Now, the polynomials $\DPx{1}$ take the role of $u(x)$ in the classical decoding up to $\dhalf$, i.e. they form the list of candidate error locator polynomials from which the decoding result is chosen. Note that the erased positions in one iteration are always named $\alpha_1$ and $\alpha_2$, there is no separate indication of the iteration. It should always be clear from the context, to which pair of positions the two variables refer.

As mentioned above, we want the degree of the polynomial $\DPx{1}$ to increase by one in each iteration. In order to ensure this degree, we force $\alpha_1$ and $\alpha_2$ to be zeros of the polynomial
\begin{equation}
\DPx{1,j+1}=a\DPx{2, j}-(x+b)\cdot\DPx{1, j}.
\label{eqn:update_wd01}
\end{equation}
Thus, we have a system of two linear equations and two unknowns, so we can give the general solution
\begin{equation}
\begin{split}
a&=\frac{\DPol{1,j}(\alpha_1)\DPol{1,j}(\alpha_2)(\alpha_1-\alpha_2)}{\DPol{1,j}(\alpha_1)\DPol{2,j}(\alpha_2)-\DPol{1,j}(\alpha_2)\DPol{2,j}(\alpha_1)},\\
b&=\frac{\DPol{1,j}(\alpha_1)\DPol{2,j}(\alpha_2)\alpha_1-\DPol{1,j}(\alpha_2)\DPol{2,j}(\alpha_1)\alpha_2}{\DPol{1,j}(\alpha_1)\DPol{2,j}(\alpha_2)-\DPol{1,j}(\alpha_2)\DPol{2,j}(\alpha_1)}.
\end{split}
\label{eqn:sol_u01}
\end{equation}
Because we do not require the polynomials to be monic, we can avoid the division in the actual implementation. However, for the analysis of the algorithm we prefer to use the formulas given as they provide the easier insight regarding the cases when the intended updating is not possible.

Of course, $\DPx{2}$ also needs to be updated to enforce the required zeros, since otherwise it will not be guaranteed in further iterations that $\DPx{1}$ still has zeros at \emph{all} positions in the corresponding erasure set $\mx_j$. 
The updating of $\DPx{2}$ is performed by
\begin{equation}
\DPx{2,j+1}=\DPx{1, j}-(a\cdot x+b)\cdot\DPx{2, j}.
\label{eqn:update_wd02}
\end{equation}
$\DPx{1, j}$ is multiplied by $1$ because we want to have $\deg \DPx{2, j+1}=\deg \DPx{1, j}$. This is derived from the fact that in \eqref{eqn:EEAauxrec}, the same auxiliary polynomial is used twice, once in the role of $\DPx{1}$ and in the next iteration in that of $\DPx{2}$. 
The correct zeros are obtained if 
\begin{equation}
\begin{split}
a&= \frac{\da\ddaa-\daa\dda}{\dda\ddaa(\alpha_1-\alpha_2)},\\
b&= \frac{\daa\dda\alpha_1-\da\ddaa\alpha_2}{\dda\ddaa(\alpha_1-\alpha_2)}.
\end{split}
\label{eqn:sol_u02}
\end{equation}
Updating according to these two rules will be called regular updating. It is directly seen, that the solutions in \eqref{eqn:sol_u01} and \eqref{eqn:sol_u02} do not always exist: Regular updating of $\DPx{1}$ is not possible if $\DPol{1,j}(\alpha_1)\DPol{2,j}(\alpha_2)-\DPol{1,j}(\alpha_2)\DPol{2,j}(\alpha_1)=0$. This happens if
\begin{equation}
\DPol{1,j}(\alpha_1)=\DPol{1,j}(\alpha_2)=0
\label{eqn:fail1}
\end{equation}
or 
\begin{equation}
\DPol{2,j}(\alpha_1)=\DPol{2,j}(\alpha_2)=0,
\label{eqn:fail2}
\end{equation}
and rarely also in other cases when the terms in the denominator of \eqref{eqn:sol_u01} are all not zero, but the denominator is. For these cases, we allow the algorithm to update $\DPx{1}$ in such a way that $\deg \DPx{1, j+1} \neq \deg \DPx{1, j}+1$. However, we keep track of this process, and perform a compensation step in some later iteration $j+j_0$ such that we get $\deg \DPx{1, j+j_0} = \deg \DPx{1, j}+j_0$. We try to choose $j_0$ as small as possible, and in most situations it is actually possible to have $j_0=2$, hence we do not intensively study the case when compensation is not immediately possible.
Additionally, if $\dda=0$ or $\ddaa=0$, the updating of $\DPx{2}$ has to be performed in a different way. Yet a closer look at the polynomials shows, that in this case it is sufficient to set $\DPx{2, j+1}=(x-\alpha_1)\DPx{2, j}$ if $\ddaa=0$ and vice versa. This might result in a polynomial of smaller degree than the intended one, but it is more important that the degree of the polynomial is not too large.

Before we study the updating procedures in the special cases indicated above, it should be noted that no further cases than the ones described before need to be distinguished. Because the auxiliary polynomials $\upolx{j-1}$ and $\upolx{j}$ fulfill the relation \cite{Sugiyama}
\begin{equation}
\upolx{j}\vpolx{j-1}-\upolx{j-1}\vpolx{j}=\pm 1,
\label{eqn:gcdeq1}
\end{equation}
where the $\vpolx{j}$ can also be calculated recursively in the EEA, but are not needed for decoding RS codes. The greatest common divisor (gcd) of two polynomials calculated in consecutive iterations of the EEA is $1$, and so this is true for $\DPx{1,0}$ and $\DPx{2, 0}$. Of course, in the further iterations, the gcd of $\DPx{1, j}$ and $\DPx{2, j}$ will at least have roots at all positions in $\mx_j$. Indeed, close examination shows that the gcd contains exactly these roots and no further common factor. 

To illustrate this with an example, we will explicitly calculate the gcd for regular updating in the first GMD iteration. For the further iterations as well as the other cases described later on, this claim can be verified in a similar manner.
First, we take a look at the proof for \eqref{eqn:gcdeq1} as given in \cite{Sugiyama}. There, the calculation of the auxiliary polynomials is written in matrix form as
\begin{equation}
\begin{pmatrix}
\upol{j} & \upol{j-1} \\
\vpol{j} & \vpol{j-1}
\end{pmatrix}
= \begin{pmatrix}\qpol{1}&1\\1&0\end{pmatrix}\cdot \begin{pmatrix}\qpol{2}&1\\1&0\end{pmatrix}\dots \begin{pmatrix}\qpol{j}&1\\1&0\end{pmatrix}
\label{eqn:proofgcold}
\end{equation}
and taking the determinant on both sides immediately gives the relation \eqref{eqn:gcdeq1}. To extend this to our approach, we also use the matrix representation, namely
\begin{equation}
\begin{pmatrix}
\DPol{1, j} & \DPol{2, j} \\
p^{(1)}(x) & p^{(2)}(x)
\end{pmatrix}= \begin{pmatrix}
\DPol{1, j-1} & \DPol{2, j-1} \\
p^{(3)}(x) & p^{(4)}(x)
\end{pmatrix}\cdot \begin{pmatrix}\qpol{j}&1\\a&\qbpol{j}\end{pmatrix},
\label{eqn:proofgc1}
\end{equation}
with polynomials $p^{(i)}(x)$ that are not used in our algorithm. For the first iteration, i.e. $j=1$, we substitute \eqref{eqn:proofgcold} for the last step of the classical decoding procedure into \eqref{eqn:proofgc1}, and by taking the determinant we obtain
\begin{equation}
\upolx{j}p^{(2)}(x)-\upolx{j-1}p^{(1)}(x)=\qpolx{1}\cdot\qbpolx{1}-a,
\label{eqn:proofgc2}
\end{equation}
the right hand side possibly multiplied by $-1$. Since we know that the gcd includes the factors $(x-\alpha_1)$ and $(x-\alpha_2)$, the gcd cannot have degree less than $2$. On the other hand, the right hand side of \eqref{eqn:proofgc2} is a polynomial of degree $2$, and it is a multiple of the gcd. Hence the gcd has degree at most $2$, so it is immediately clear that the gcd consists of exactly the two factors given, in particular no root at any value $\alpha_i$ that is to be erased in a later iteration, is contained in the gcd.

Now we will turn to the special cases, where the updating rule \eqref{eqn:update_wd01} cannot be used. First, we study the case given in \eqref{eqn:fail2}. Due to the fact that $\DPx{1}$ and $\DPx{2}$ are coprime, we have $\DPol{1,j}(\alpha_1)\neq 0, \DPol{1,j}(\alpha_2)\neq 0$. In such a case, $(x+b)\DPx{1, j}$ can only have a root at one of the required zeros, and adding a multiple of $\DPx{2, j}$ cannot bring either position to zero. Therefore, we choose to set $\deg \DPx{1, j+1}=\deg \DPx{1, j}+2$. Then it is easy to find the updating rules
\begin{equation}
\begin{split}
\DPx{1, j+1} &= (x-\alpha_1)(x-\alpha_2)\DPx{1, j}, \\
\DPx{2, j+1} &= \DPx{2, j}.
\end{split}
\label{eqn:update_wd+1}
\end{equation}
In the next iteration, the decoder should try to compensate for this decision. The thought leading to the result is the following: If we check how the polynomial $\DPx{1, j+2}$ is composed of $\DPx{1, j}$ and $\DPx{2, j}$ for regular updating, we find that the first is multiplied by a polynomial of degree $2$ and the latter by a polynomial of degree $1$. We therefore set
\begin{equation}
\begin{split}
\DPx{1, j+2} &= (x+a)\DPx{2, j+1}+b\DPx{1, j+1}, \\
\DPx{2, j+2} &= (x-\alpha_1)(x-\alpha_2)\DPx{2, j+1},
\end{split}
\label{eqn:comp_wd+1}
\end{equation}
which in combination with \eqref{eqn:update_wd+1} gives the desired degrees for two updating steps.
$\DPx{1, j+2}$ has the desired zeros if 
\begin{equation}
\begin{split}
a&=\frac{\DPol{1}(\alpha_1)\DPol{2}(\alpha_2)\alpha_1-\DPol{1}(\alpha_2)\DPol{2}(\alpha_1)\alpha_2}{\DPol{1}(\alpha_1)\DPol{2}(\alpha_2)-\DPol{1}(\alpha_2)\DPol{2}(\alpha_1)},\\
b&=\frac{\DPol{2}(\alpha_1)\DPol{2}(\alpha_2)(\alpha_1-\alpha_2)}{\DPol{1}(\alpha_1)\DPol{2}(\alpha_2)-\DPol{1}(\alpha_2)\DPol{2}(\alpha_1)};
\end{split}
\label{eqn:comp_wd+12}
\end{equation}
we abbreviated $\DPol{1}=\DPol{1, j+1}$ and $\DPol{2}=\DPol{2, j+1}$.

If $\DPol{1,j}(\alpha_1)\DPol{2,j}(\alpha_2)-\DPol{1,j}(\alpha_2)\DPol{2,j}(\alpha_1)=0$ without any of the involved terms being zero, we perform the update of $\DPx{1}$ as in \eqref{eqn:update_wd+1}. However, the updating of $\DPx{2}$ still is done according to \eqref{eqn:update_wd02}, since (of course) $\DPx{2, j}$ does not yet have the required zeros. After the compensation step, $\DPx{2}$ will then have the same degree as $\DPx{1}$. However, recursive substitution, in order to get $\DPx{1}$ in dependence of $\DPx{1, 0}$ and $\DPx{2, 0}$ shows that these polynomials are multiplied by the intended degree in $\DPx{1}$, so this fact does not cause major problems, and simulations have shown that correct decoding is actually achieved with this setup.

The last special case that needs to be taken into account is $\DPol{1,j}(\alpha_1)=\DPol{1,j}(\alpha_2)=0$. Here, there is no need to update $\DPx{1}$, but forcing the zeros in $\DPx{2, j+1}$ is not possible with the formula given in \eqref{eqn:update_wd02}: The solution in \eqref{eqn:sol_u02} is valid, but the fact that $a=b=0$ implies that $\DPx{2, j}$ is discarded, hence all further solutions would be multiples of $\DPx{1, j}$, which is not wanted. Therefore, we use the updating rules
\begin{equation}
\begin{split}
\DPx{1, j+1} &= \DPx{1, j}, \\
\DPx{2, j+1} &= (x-\alpha_1)(x-\alpha_2)\DPx{2, j}.
\end{split}
\label{eqn:update_wd-1}
\end{equation}

The compensation step then is given as
\begin{equation}
\begin{split}
\DPx{1, j+2} &= (x-\alpha_1)(x-\alpha_2)\DPx{1, j+1}, \\
\DPx{2, j+2} &= a\DPx{2, j+1}-(x+b)\DPx{1, j+1},
\end{split}
\label{eqn:comp_wd-1}
\end{equation}
where
\begin{equation}
\begin{split}
a&= \frac{\da\daa(\alpha_1-\alpha_2)}{\daa\dda-\da\ddaa} \\
b&= \frac{\da\ddaa\alpha_1-\daa\dda\alpha_2}{\daa\dda-\da\ddaa}.
\end{split}
\label{eqn:comp_wd-12}
\end{equation}
Performing the compensation steps, both that in \eqref{eqn:comp_wd+1} and in \eqref{eqn:comp_wd-1} isn't always possible either. The situations in which compensation fails are actually the same as those where regular updating fails. Consequently, the same special updating rules are used again in this cases. If the same rule is used twice, two compensation steps are required later on. On the other hand, \eqref{eqn:update_wd+1} and \eqref{eqn:update_wd-1} serve as compensation steps for each other. Further, compensation is possible but should not be performed in \eqref{eqn:comp_wd+1} if $\DPol{2, j+1}(\alpha_1)$ or $\DPol{2, j+1}(\alpha_2)$ is zero: In this case, by performing the compensation step, one discards the polynomial $\DPx{1, j+1}$, and all error locator polynomials obtained further are multiples of $\DPx{2, j+1}$ which is not wanted. Therefore, it is better to perform regular updating in this situation and try compensation in the next iteration. The same holds if the compensation in \eqref{eqn:comp_wd-1} is to be performed and $\DPol{1, j+1}(\alpha_1)=0$ or $\DPol{1, j+1}(\alpha_2)=0$. 

The situation in these special cases is a little different if $\qhpolx{1}$ exists. As mentioned before, the updating can be written to avoid the division in the calculation of the coefficients - such a case is equivalent to multiplying both $\DPx{1, j}$ and $\DPx{2, j}$ by a constant. As a result, $\DPx{1, j+1}=c\DPx{2, j+1}$ with constant $c$ would be obtained, and in the next iteration, forcing more zeros would result in $\DPx{1, j+2}=0$. This is still true for the changed definition of $\DPx{2, 0}$, and so updating is best done as presented before. However, because we used the term $\qhpolx{1}\upolx{j}$ in the definition of $\DPx{2, 0}$, in these cases another solution may be obtained: As long as the constant multiplied to $\DPx{2, j}$ is not zero, the result includes a term $a_1\cdot x^i\DPx{1, j}$ with $i > 1$. Therefore, in such cases an additional solution - of the form $\DPx{1, j+1}=a\DPx{2, j}+b\DPx{1, j}$ - is a valid solution and therefore it is stored, increasing the maximum list size at the decoder output.

We conclude this section by sketching how the formulas given for the GMD extension interact as an algorithm. We set $\bar{\Delta}:=\da\ddaa-\daa\dda$ to obtain a shorter notation. The variable $dd$ introduced in the algorithm is used to keep track of the special updatings performed.

\begin{algorithm}[H]
\SetLine
\dontprintsemicolon
\KwIn{Polynomials $\DPx{1, 0}$, $\DPx{2, 0}$, \\erasure sets $\mx_j$}
\KwOut{List $\mathcal{L}$ of candidate error locators}
Initialization: $j=0$, 
$dd=0$, $\mathcal{L}=\{\DPx{1, 0}\}$\;
\While{$\deg \DPx{1, j} < d-2$}
{
	calculate $\bar{\Delta}$ from $\mx_{j+1}\setminus\mx_{j}$\;
	\uIf{$dd=0$ and $\bar{\Delta} \neq 0$}
		{update according to \eqref{eqn:update_wd01} and \eqref{eqn:update_wd02}}
	\uElseIf{$\bar{\Delta} = 0$}
		{perform special updating and adjust $dd$:\;
		\eqref{eqn:update_wd+1} $\Rightarrow \; dd = dd+1$\;
		\eqref{eqn:update_wd-1} $\Rightarrow \; dd = dd-1$\;
		}
	\uElseIf{$dd > 0$ (and $\bar{\Delta} \neq 0$)}
		{perform compensation step \eqref{eqn:comp_wd+1}, $dd=dd-1$\;}
	\ElseIf{$dd < 0$ (and $\bar{\Delta} \neq 0$)}
		{perform compensation step \eqref{eqn:comp_wd-1}, $dd=dd+1$\;}
	store $\DPx{1, j+1}$ in $\mathcal{L}$\;
	$j=j+1$\;
}
\caption{GMD extension}
\end{algorithm}

\section{Selection of the Best Solution}\label{sec:select}
In this section, we shortly discuss how to select the best solution from the list of candidate error locators. Although the distance criteria used in most cases are trivial, the straightforward approach requires to perform error evaluation for all $\order(d)$ candidate error locators, with a complexity of $\order(n^2)$ each. Hence, in this straightforward approach the overall complexity is $\order(n^3)$ and determined by the evaluation step. 

The method suggested by K\"otter \cite{Koetter_GMD} is to use evaluation vectors instead of polynomials during the algorithm. Instead of using the polynomials, evaluation vectors are calculated in the initialization to the GMD extension. In these vectors, every component corresponds to the evaluation of the polynomial at a certain field value, i.e. we substitute
\begin{equation}
\DPx{1, 0} \leftrightarrow \left[ \DPol{1, 0}(1), \DPol{1, 0}(\alpha), \dots \DPol{1, 0}(\alpha^{n-1})\right].
\end{equation}
and so on. Consequently, polynomial multiplications are replaced by elementwise vector multiplications, e.g. 
\begin{equation}
x\cdot p(x) \leftrightarrow \left[p(1), \alpha p(\alpha), \alpha^2 p(\alpha^2), \dots, \alpha^{n-1} p(\alpha^{n-1})\right].
\label{eqn:xvector}
\end{equation}
The evaluation of a polynomial is now simply the extraction of one component from a vector. It can be verified that the complexity of the main step, namely finding all candidate error locators, can still be performed with a complexity of $\order(n^2)$. Further, the selection of the best solution is now also possible with complexity $\order(n^2)$: Using a weighted hamming metric, it is only important which positions are in error, while the actual error value is not important. Since these positions can be easily extracted from the evaluation vectors of the candidate polynomials - the positions where the evaluation vectors are zero - the complexity of calculating the weighted Hamming weight of a single candidate is therefore only $\order(n)$, and so finding the best solution among $\order(d)$ candidates can be done with complexity $\order(n^2)$.

\section{Summary and Conclusion}\label{sec:summary}
We presented a method that is capable of performing Generalized Minimum Distance decoding of RS codes with an overall complexity of $\order(n^2)$. A method that exhibits the same performance had already been introduced by K\"otter in 1996 \cite{Koetter_GMD}. This is not surprising, since K\"otters algorithm extends the Berlekamp-Massey algorithm, and ours the extended Euclidean algorithm, and these two algorithms are known to be equivalent for decoding up to half the minimum distance. However, having a different problem formulation and erasing strategy, we do not always have the same intermediate results. 

\bibliographystyle{IEEEtran}
\bibliography{bib_eea_ext}

\end{document}